\documentstyle[aps,prb]{revtex}
\renewcommand {\theequation}{\arabic{section}.\arabic{equation}}

\newcounter {pzz}
\newcommand{\pppp}{\setcounter{pzz}{\value{equation}}\setcounter{equation}{0}\renewcommand {\theequation}{\arabic{section}.\arabic{equation}} }

\draft

\begin{document}
\title{Spin polaron damping in the spin-fermion model for cuprate
superconductors}
 
\author{R.O.\ Kuzian}
\address{Institute for Materials Science, \\
Krjijanovskogo 3, \\
252180 Kiev, Ukraine}
\author{R.\ Hayn}
\address{Max Planck Arbeitsgruppe Elektronensysteme,\\
Technische Universit\"at Dresden,\\
D-01062\\
Dresden, Germany}
\author{A.F.\ Barabanov}
\address{Institute for High Pressure Physics, \\
142092 Troitsk, Moscow region, Russia}
\author{L.A. Maksimov}
\address{Russian Research Center Kurchatov Institute, \\
Kurchatov sq.46, \\
123182 Moscow, Russia}
\date{\today }
\maketitle

\begin{abstract}
A self-consistent, spin rotational invariant Green's function procedure has
been developed to calculate the spectral function of carrier excitations in
the spin-fermion model for the CuO$_2$ plane. We start from the mean field
description of a spin polaron in the Mori-Zwanzig projection method. In
order to determine the spin polaron lifetime in the self-consistent Born
approximation, the self-energy is expressed by an irreducible Green's
function. Both, spin polaron and bare hole spectral functions are
calculated. The numerical results show a well pronounced quasiparticle peak
near the bottom of the dispersion at $(\pi /2,\pi /2)$, the absence of the
quasiparticle at the $\Gamma $-point, a rather large damping away from the
minimum and an asymmetry of the spectral function with respect to the
antiferromagnetic Brillouin zone. These findings are in qualitative
agreement with photoemission data for undoped cuprates. The direct
oxygen-oxygen hopping is responsible for a more isotropic minimum at $(\pi
/2,\pi /2)$.
\end{abstract}


%\pacs{71.27.+a,71.10.Fd,75.30.Mb}
%PACS numbers: 74.25.Jb, 74.72.-h,75.10.-b

%\newpage

\pagestyle{plain} \setcounter{page}{1}

\section{Introduction}

In recent years there was some progress to answer the important question on
the nature of quasiparticles in cuprate superconductors, but the task is not
yet completed. Usually, the problem is reduced to that of the hole motion in
a two dimensional quantum antiferromagnet. Many work was devoted to the $t$-$
J$ model \cite{dagotto,schmitt,kane,eder,martinez} treating the
antiferromagnet as a two-sublattice N\'eel state. It was shown that the hole
motion is characterized by a quasiparticle band of bandwidth $\sim 2J$ that
is separated from the incoherent part. \cite
{dagotto,schmitt,kane,eder,martinez} The minimum of the dispersion is found
at the momentum $(\pi /2,\pi /2)$ (lattice constant $g=1$) nearly degenerate
with the energy at $(\pi ,0)$. Such a result was obtained by several
methods. One of them is the self-consistent Born approximation (SCBA) for a
reduced version of the $t$-$J$ model consisting of spinless holes coupled to
spinons. \cite{schmitt,kane,martinez,liu,plakida} That method became
especially popular since it is sufficiently simple and allows the
calculation of the spectral function. It gives rise to a dispersion and also
a spectral function which is symmetric with respect to the magnetic
Brillouin zone (BZ). Similar results were obtained for the Emery model. \cite
{eder2,kabanov,starykh}

In a remarkable experiment \cite{wells} the quasiparticle dispersion of such
a spin polaron was measured directly by creating a photohole in the
antiferromagnetic insulator Sr$_2$CuO$_2$Cl$_2$. Whereas the width of the
dispersion was in reasonable agreement with the theoretical predictions, its
shape had some deviations especially at the point $(\pi ,0)$ in momentum
space. Afterwards it was shown that the deviations can be diminished by
taking into account hopping terms to second and third neighbors in the $t$-$J
$ model. \cite{nazarenko,bala,yushankhai} These additional terms
correspond to a proper treatment of direct oxygen-oxygen hopping in the
Emery model. \cite{starykh,yushankhai} However, some deviations between
theory and experiment remain. So, the measurement shows no quasiparticle
weight at the $\Gamma $-point $(0,0)$ but the spin polaron in the $t$-$J$
model has. Next, one observes quite a remarkable damping especially away
from the minimum. Some theories (especially the usual SCBA) are symmetric
with respect to the antiferromagnetic (AFM) BZ, but the experimental data
show a strong asymmetry: after crossing the AFM BZ boundary (coming from the
$\Gamma $-point) the quasiparticle looses its weight nearly abruptly. That
deficiency, however, seems to be merely a property of the usual SCBA and not
of the $t$-$J$ model itself. \cite{eder3,eskes}

In the usual SCBA there is no dispersion of the bare spinless hole; the
dispersion appears only due to the coupling to the spinons. Therefore, one
might expect large vertex corrections. In fact, however, the low order
vertex corrections cancel in the pure $t$-$J$ model as it was noted in
Refs.\ \onlinecite{martinez,liu}. Such a cancelation is absent in the more
realistic Emery model. But there are many indications (see also Ref.\
\onlinecite{vos}), and we will add some arguments below, that the
photoemission data cannot be explained by the $t$-$J$ model alone, at least
not throughout the whole BZ.

We concentrate in the present paper on the spin-fermion model (the reduced
form of the three-band Emery model) \cite{em,emrei} for the CuO$_2$ plane
and present a new scheme to calculate the spectral function starting from
the mean field description \cite{bkm1,bkm2} of the spin polaron. One can
expect that the fluctuations around the mean field spin polaron are smaller
than around the dispersionless spinless hole. The mean field description of
the spin polaron in the Emery model was introduced in Refs.\
\onlinecite{bkm1,bkm2,barVan}. It is based on a description of the hole
motion not in the two-sublattice N\'eel state but in the spherically
symmetric (in spin space) AFM state with long range order (LRO). Such a spin
singlet state is characterized here by a vanishing expectation value of the
magnetization $\left\langle S_i^z\right\rangle =0$ (the only possible
nonzero expectation values are scalars with respect to rotations in spin
space), but a finite value of the spin-spin correlations at infinite
distance $\left\langle \vec S_0\vec S _R\right\rangle \to (-1)^RM$ for $R\to
\infty $. It can be constructed, for instance, from a resonating valence
bond (RVB) state with long-range bonds that decay not too fast. \cite{liang}
The spherically symmetric state is an appealing candidate to study the
transition from the paramagnetic spin liquid (with only short range
correlations) to a state with long-range order.

A spherically symmetric Green's function (GF) theory for Heisenberg models
was constructed by Kondo and Yamaji (KY) \cite{kondo} and it will be used as
one ingredient of the present theory. It is a GF decoupling scheme which is
applicable to cases in which there is no symmetry breaking, especially to
low dimensional systems (for the 2D square lattice see Refs.\ \onlinecite
{shimahara,barabanov94}). As a result, the carrier excitations (moving in
the spherically symmetric background) preserve the periodicity relative to
the full BZ (not the magnetic one). Namely this leads to a qualitative
agreement with the experimental results in the undoped compound Sr$_2$CuO$_2$
Cl$_2$. An explanation might be that the actual experiment was done at room
temperature, whereas the N\'eel temperature of Sr$_2$CuO$_2$Cl$_2$ is only
250 K.

The aim of the present work is twofold. In the first place it is a
methodical work proposing a new technique to calculate the spectral function
of quasiparticles with complicated operator structure. Our technique
combines the Mori-Zwanzig projection method with a self-consistent
calculation of the self-energy in terms of irreducible GF (Tserkovnikov
technique \cite{tserkovnikov}). In difference to our recent works \cite
{barabanov97,barjetp97} we introduce a simultaneous description of a spin
polaron and a bare hole, and investigate the effects of direct oxygen-oxygen
hopping which only gives us the possibility to reach a realistic
description. Furthermore, we discuss also the upper parts of the spectrum in
the spin-fermion model whose lowest branch corresponds to the spin polaron.
The second aim is to present numerical results for the spectral function. We
will distinguish three regions in ${\bf k}$-space where the spin polaron
exists as a quasiparticle with infinite lifetime, with finite lifetime or
where it is strongly overdamped. Our method leads to several, qualitatively
new features which are not present in the SCBA of spinless holes coupled to
spinons. But these features may be observed in the experimental data for Sr$
_2$CuO$_2$Cl$_2$. We are concentrating here on a qualitative discussion
rather than fine tuning the parameters of the Emery model in its
spin-fermion form to obtain maximal agreement with experiment.

Our paper is organized as follows: after presenting the set of basis
operators which define the mean field description of the spin polaron (Sec.\
3) we show the general procedure to calculate the damping (Sec.\ 4) and
calculate the corresponding vertex (Sec.\ 5). The self-consistent set of
equations for the GF is given in Sec.\ 6 and the results are presented in
Sec.\ 7.
%In the Appendix A we give the derivation of Tserkovnikov expression for
%the self-energy (\ref{Rirr}) and compare it briefly with the conventional
%projection-technique expression,
%and in the Appendix B we give a short review the Kondo-Yamaji theory
%\cite{kondo} for the isotropic Heisenberg
%spin-half antiferromagnet on a square lattice.

\section{Spin-fermion model}
\pppp

An extra hole propagating in the ${\rm CuO}_2$ plane of cuprates strongly
interacts with the antiferromagnetically correlated copper spin subsystem.
The main features of the hole motion are described by the model: \cite
{em,emrei}

\begin{equation}
\label{ham}\hat H=\hat \tau +\hat J+\hat h\;,
\end{equation}
\begin{equation}
\label{J}\hat \tau =4\tau \sum_{{\bf R}}p_{{\bf R}}^{\dagger }\left( \frac 12
+\tilde S_{{\bf R}}\right) p_{{\bf R}}\;,\qquad \hat J=\frac J2\sum_{{\bf R},
{\bf g}}S_{{\bf R}}^\alpha S_{{\bf R}+{\bf g}}^\alpha \;,
\end{equation}
\begin{equation}
\label{hh}\hat h=-h\sum_{{\bf R}}\left[ c_{{\bf R+a}_x}^{\dagger }\left( c_{
{\bf R+a}_y}+c_{{\bf R-a}_y}+c_{{\bf R+g}_x{\bf +a}_y}+c_{{\bf R+g}_x-{\bf a}
_y}\right) +h.c.\right] \;,
\end{equation}
with
\begin{equation}
p_{{\bf R}}=\frac 12\sum_{{\bf a}}c_{{\bf R}+{\bf a}},\quad \tilde S_{{\bf R}
}\equiv S_{{\bf R}}^\alpha \hat \sigma ^\alpha ,\quad \left\{ p_{{\bf R}},p_{
{\bf R}^{\prime }}^{\dagger }\right\} =\delta _{{\bf R,R}^{\prime }}+\frac 14
\sum_{{\bf g}}\delta _{{\bf R,R}^{\prime }+{\bf g}}\;,
\end{equation}
where ${\bf a}_{x,y}=\frac 12{\bf g}_{x,y},{\bf g}=\pm {\bf g}_x,\pm {\bf g}
_y$.

Here and below the summation over repeated indexes is understood everywhere;
$\{\ldots ,\ldots \},\ [\ldots ,\ldots ]$ stand for anticommutator and
commutator respectively; ${\bf g}_{x,y}$ are basis vectors of a copper
square lattice ( $|{\bf g}|\equiv 1$), ${\bf R}+{\bf a}$ are four vectors of
O sites nearest to the Cu site ${\bf R}$; the operator $c_{{\bf R+a}
}^{\dagger }$ creates a hole predominantly at the O site (the spin index is
dropped in order to simplify the notations); $\hat \sigma ^\alpha $ are the
Pauli matrices; the operator ${\bf S}$ represents the localized spin on the
copper site. We do not introduce the explicit relative phases of $p$- and $d$
-orbitals since they can be transformed out by redefining the operators with
phase factors $\exp \left( \imath {\bf q}_0{\bf R}\right) ,{\bf q}_0=\left(
\pi ,\pi \right) $. In order to compare our results with other authors and
with experiment we restore these phases by changing in the final results $
{\bf k} \rightarrow {\bf k}^{\prime }={\bf k}-{\bf q}_0$. The parameter $
\tau $ is the hopping amplitude of oxygen holes that take into account the
coupling of the hole motion with the copper spin subsystem, $J$ is the
constant of nearest neighbor AFM exchange between the copper spins. For the
parameter values we take $\tau \sim$ 0.5eV, $J=0.2\tau$ throughout the
paper, $h=0$, or $h=0.3\tau $ to illustrate the influence of direct
oxygen-oxygen hopping.

The Hamiltonian (\ref{ham}) corresponds to the regime $U_d\gg \left|
\varepsilon _p-\varepsilon _d\right| \gg t_{pd}$ in the Emery model. It can
be obtained \cite{zaanen} e.g.\ by a canonical transformation of bare
fermion operators $\bar a ={\rm e}^{-S}a{\rm e}^S$. Then the transformed
operators $a$ which enter the Hamiltonian (\ref{ham}) acquire the admixture
of bare operators at adjacent sites. For example, the ''oxygen hole''
operator in the Eq.\ (\ref{hh}) may be expressed as
\begin{equation}
\label{sw}c_{{\bf R+a,s}}={\rm e}^{\bar S}\bar c_{{\bf R+a,s}}{\rm e}^{-\bar
S}\approx \bar c_{{\bf R+a,s}}+\frac{t_{pd}}{\left| \varepsilon
_p-\varepsilon _d\right| }\bar d_{{\bf R,}s}\left( 1-\bar d_{{\bf R,}
-s}^{\dagger }\bar d_{{\bf R,}-s}\right) .
\end{equation}
This operator annihilates the hole in a state that is an antibonding (in
hole notation) combination of the oxygen $p_\sigma $ state and the singly
occupied copper $d_{x^2-y^2}$ state at adjacent sites.

The non-bosonic character of spin operators and the complexity of the spin
subsystem ground state $\left| {\cal G}\right\rangle $ make a perturbative
treatment of the Hamiltonian (\ref{ham}) difficult. For this reason we
develop in the next sections the method of Green's function calculation
which combine the Mori-Zwanzig projection method (PM) \cite{mori,FuldeB}
with the Tserkovnikov technique. The PM provides the possibility to allow
for the local correlations thoroughly. For example, the local constraints
are naturally included in the calculation. We thus obtain the proper
description of small polaron (so called Zhang-Rice (ZR) singlet) \cite
{bkm1,ZhR} formation that is governed by the Kondo term in $\hat \tau $. On
the other hand, the PM always provides the Green's function as the sum of
simple poles, since only a finite number of basis operators may be included
in the calculation. The consideration of polaron scattering on spin
excitations, which is responsible for the damping of the polaron state,
demands an infinite extension of the basis set. This can be done effectively
by the self-consistent Born approximation (SCBA) for the polaron
self-energy, calculated within the Tserkovnikov technique. \cite
{tserkovnikov} That means that we suppose the effect of the residual
polaron-spin interaction to be rather small and tractable within a
self-consistent perturbative scheme. The main point here is the appropriate
choice of the operator basis set for PM and a careful division of the
interaction into mean field and residual part.

\section{Basis operators and projection method}
\pppp

>From the very beginning we want to take into account properly the local
correlations $\hat \tau $ without the violation of spin commutation
relations. We consider a spin-liquid state with spin rotational symmetry for
the copper subsystem, e.g.\ the spin-spin correlation functions satisfy the
relation $C_{{\bf r}}\equiv \left\langle S_{{\bf R}}^\alpha S_{{\bf R}+{\bf r
}}^\alpha \right\rangle =3\left\langle S_{{\bf R}}^{x\left( y,z\right) }S_{
{\bf R}+{\bf r}}^{x\left( y,z\right) }\right\rangle $. We choose a set of
three basis operators. The first two of them constitute the ZR polaron
\begin{equation}
\label{b1}B_{1,{\bf k}}=\frac 1{\beta _{{\bf k}}\sqrt{N}}\sum_{{\bf R}}{\rm e
}^{-\imath {\bf kR}}p_{{\bf R}}\;,\quad B_{2,{\bf k}}=\frac 1{\nu _{{\bf k}}
\sqrt{N}}\sum_{{\bf R}}{\rm e}^{-\imath {\bf kR}}\tilde S_{{\bf R}}p_{{\bf R}
}\;,
\end{equation}
where the factors $\beta _{{\bf k}}$ and $\nu _{{\bf k}}$ arise due to the
orthonormalization
\begin{equation}
\beta _{{\bf k}}=\sqrt{1+\gamma _{{\bf k}}}\;,\quad \nu _{{\bf k}}=\sqrt{
\frac 34+C_{{\bf g}}\gamma _{{\bf k}}}\;,
\end{equation}
and $\gamma _{{\bf k}}\equiv \frac 12\left( \cos k_x+\cos k_y\right) $. In
fact, the basis operators (\ref{b1}) can be combined in two ways
corresponding to the ZR singlet and part of the ZR triplet state. Only the
lower lying singlet combination builds the spin polaron quasiparticle. We
can also write
\begin{equation}
B_{1,{\bf k}}=\frac 1{\beta _{{\bf k}}}\left[ \cos \left( \frac{k_x}2\right)
c_{{\bf k,}x}+\cos \left( \frac{k_y}2\right) c_{{\bf k,}y}\right]
\end{equation}
where
$$
c_{{\bf k,}x,y}=\frac 1{\sqrt{N}}\sum_{{\bf R}}{\rm e}^{-\imath {\bf k}
\left( {\bf R+a}_{x,y}\right) }c_{{\bf R+a}_{x,y}}.
$$
Now it is easy to see that one has to introduce the operator
\begin{equation}
\label{b3}B_{3,{\bf k}}=\frac 1{\beta _{{\bf k}}}\left[ \cos \left( \frac{k_y
}2\right) c_{{\bf k,}x}-\cos \left( \frac{k_x}2\right) c_{{\bf k,}y}\right]
\end{equation}
in order to represent the full set of bare hole operators. That is important
if we want to find the hole spectral weight of the spin polaron band.

In a first step we construct the three eigenoperators ${\cal B}_{i,{\bf k}}$
and the corresponding bands $\Omega _{{\bf k}}^{(i)}$ in the mean field
approach:
\begin{equation}
\label{eio}{\cal B}_{i,{\bf k}}=\alpha _j^{(i)}\left( {\bf k}\right) B_{j,
{\bf k}}\;,\quad {\cal B}_{{\bf k}}\equiv {\cal B}_{1,{\bf k}}\;,\quad
\Omega _{{\bf k}}\equiv \Omega _{{\bf k}}^{\left( 1\right) }\;,
\end{equation}
where we introduced a simplified notation for the most interesting lowest
spin polaron band $\Omega _{{\bf k}}$ with the polaron annihilation
eigenoperator ${\cal B}_{{\bf k}}$. We use the projection method of Mori and
Zwanzig \cite{mori,FuldeB} and introduce the retarded Green's functions
(GF):
\begin{equation}
\label{gfd}G_{ij}({\bf k},\omega )=\langle B_{i,{\bf k}}|B_{j,{\bf k}
}^{\dagger }\rangle _\omega \equiv -\imath \int_{t^{\prime }}^\infty
\!\!dte^{\imath \omega (t-t^{\prime })}\langle \{B_{i,{\bf k}}(t),B_{j,{\bf k
}}^{\dagger }(t^{\prime })\}\rangle \;.
\end{equation}
Here we use Zubarev's notations. \cite{zubarev} Within the PM the equations
of motion for the GF (\ref{gfd}) are projected onto the subspace spanned by
the operators $B_{i,{\bf k}}$ which leads to the following eigenvalue
problem to determine $\alpha _j^{\left( n\right) }$ and $\Omega _{{\bf k}
}^{\left( n\right) }$:
\begin{equation}
\label{mf}\omega G_{ij}^{(0)}({\bf k},\omega )=\delta _{ij}+{\cal L}
_{il}G_{lj}^{(0)}({\bf k},\omega ),\quad \left( {\cal L}_{ij}-\Omega _{{\bf k
}}^{\left( n\right) }\delta _{ij}\right) \alpha _j^{\left( n\right) }\left(
{\bf k}\right) =0
\end{equation}
where
\begin{equation}
\label{Liu}{\cal L}_{ij}\equiv \left\langle \left\{ \left[ B_{i,{\bf k}},
\hat H\right] ,B_{j,{\bf k}}^{\dagger }\right\} \right\rangle ;\quad
\left\langle \left\{ B_{i,{\bf k}},B_{j,{\bf k}}^{\dagger }\right\}
\right\rangle =\delta _{ij}\;.
\end{equation}
An explicit calculation expresses ${\cal L}_{ij}$ in terms of the two-site
spin-spin correlation functions $C_{{\bf r}}$ and gives:
\begin{eqnarray}
{\cal L}_{11}&=&2\tau \beta _{{\bf k}}^2-\frac{8h\pi _{{\bf k}}^2}{\beta _{
{\bf k}}^2} \; ,
\quad
{\cal L}_{12}={\cal L}_{21}=4\tau \beta _{{\bf k}}\nu _{{\bf k}} \; ,
\quad
{\cal L}_{13}={\cal L}_{31}=-\frac{4h\pi _{{\bf k}}\delta _{{\bf k}}}{\beta
_{{\bf k}}^2},
\nonumber \\
{\cal L}_{22}&=&\frac \tau {\nu _{{\bf k}}^2}\left[ -\frac 98+C_{{\bf g}}\left(
1-4\gamma _{{\bf k}}\right) +\frac 18\sum_{{\bf g}_1\neq {\bf g}_2}{\rm e}
^{-\imath {\bf k}\left( {\bf g}_1-{\bf g}_2\right) }C_{{\bf g}_1-{\bf g}
_2}\right]
\nonumber \\
&& +\frac J{\nu _{{\bf k}}^2}C_{{\bf g}}\left( \gamma _{{\bf k}}-4\right) -
\frac h{\nu _{{\bf k}}^2}\left[ \frac 32+4C_{{\bf g}}\gamma _{{\bf
k}}+2C_{{\bf g}
_x+{\bf g}_y}\left( \gamma _{{\bf k}}^2-\delta _{{\bf k}}^2\right) \right] ,
\nonumber \\
{\cal L}_{23}&=&{\cal L}_{32}=0 \; ,
\quad
{\cal L}_{33}=\frac{8h\pi _{{\bf k}}^2}{\beta _{{\bf k}}^2}
\end{eqnarray}
where
$$
\delta _{{\bf k}}\equiv \frac 12\left( -\cos k_x+\cos k_y\right) ,\quad \pi
_{{\bf k}}\equiv \cos \left( \frac{k_x}2\right) \cos \left( \frac{k_y}2
\right) .
$$

The projected equation of motion (\ref{mf}) defines the mean field Green's
functions $G_{ij}^{(0)}$ with three bands $\Omega_{{\bf k}}
= \Omega_{{\bf k}}^{(1)},\ldots
,\Omega _{{\bf k}}^{(3)}$ corresponding to the ZR singlet, the nonbonding
oxygen and the triplet bands. The mean field spectrum of all three bands is
shown in Figs.\ 1(a) and 1(b) for two different parameter sets. In the
following we will mainly concentrate on the eigenoperator of the lowest
band, the polaron eigenoperator ${\cal B}_{{\bf k}}={\cal B}_{1,{\bf k}}$.
The GF of the polaron quasiparticle is defined as:
\begin{equation}
\label{pGf}G_p({\bf k},\omega )=\langle {\cal B}_{{\bf k}}|{\cal B}_{{\bf k}
}^{\dagger }\rangle _\omega \;.
\end{equation}
One has to distinguish between the polaron GF defined in (\ref{pGf}) and the
GF $G_h({\bf k},\omega )$ which gives the number of holes in a unit cell and
is introduced as:
\begin{equation}
\label{Ghh}G_h({\bf k},\omega )=\langle c_{{\bf k,}x}|c_{{\bf k,}x}^{\dagger
}\rangle _\omega +\langle c_{{\bf k,}y}|c_{{\bf k,}y}^{\dagger }\rangle
_\omega =\langle B_{1,{\bf k}}|B_{1,{\bf k}}^{\dagger }\rangle _\omega
+\langle B_{3,{\bf k}}|B_{3,{\bf k}}^{\dagger }\rangle _\omega \;.
\end{equation}
Please note that according to (\ref{sw}) the GF $G_h$ counts not only the
holes at oxygen sites, but has a contribution at the copper sites also.
Below we shall suppose that the intensity given by the photoemission
experiments can be roughly compared with the hole spectral function which is
dictated by $G_h({\bf k},\omega )$. In the mean field approximation $
G_h^{(0)}({\bf k},\omega )$ has the form
\begin{equation}
\label{Ghp}G_h^{(0)}({\bf k},\omega )=\sum_i\frac{\left| \alpha _1^{\left(
i\right) }\left( {\bf k}\right) \right| ^2+\left| \alpha _3^{\left( i\right)
}\left( {\bf k}\right) \right| ^2}{\omega -\Omega _{{\bf k}}^{\left(
i\right) }}=\sum_i\frac{Z^{(i)}\left( {\bf k}\right) }{\omega -\Omega _{{\bf
k}}^{\left( i\right) }}\;,
\end{equation}
which defines the pole strength $Z^{(i)}\left( {\bf k}\right) $ of all three
bands. If we restrict ourselves to the vicinity of the lowest pole $\Omega _{
{\bf k}}=\Omega _{{\bf k}}^{(1)}$ the hole GF can be approximated by
\begin{equation}
\label{lowpo}G_h^{(0)}({\bf k},\omega )\approx {Z\left( {\bf k}\right) }
G_p^{(0)}({\bf k},\omega )\;,
\end{equation}
with
\begin{equation}
\label{lowpo2}Z\left( {\bf k}\right) =Z^{(1)}\left( {\bf k}\right) \;,\quad
\mbox{and}\quad G_p^{(0)}=\frac 1{\omega -\Omega _{{\bf k}}}\;.
\end{equation}

Let us first discuss the mean field spectrum of all three bands for the two
parameter sets with and without direct oxygen-oxygen hopping $h$ (Figs.\
1(a) and 1(b)). First of all, one observes a well separated singlet band
between 1 and 2 eV below the nonbonding one. The singlet-triplet splitting
is maximal at $(\pi,\pi)$ and roughly 3 eV (assuming $\tau$ to be 0.5 eV).
That agrees quite reasonably with the splitting between the ZR singlet and
triplet states of 3.4 eV that were found in calculations for a CuO$_4$
cluster. \cite{eskes2} And experimental indications for the triplet state
were found in a recent photoemission measurement of the substance Ba$_2$Cu$_3
$O$_4$Cl$_2$. \cite{schmelz} It is also interesting to note that the singlet
band dispersion obeys nearly the antiferromagnetic symmetry (with only small
deviations due to the spin rotational invariant ground state) but the
triplet band not. The nonbonding band receives a dispersion due to $h$. It
has the maximum (which would correspond to a binding energy maximum in a
spectroscopic measurement) at $(\pi,\pi)$. One should note, however, that we
did not include into our model all the three oxygen 2$p$ orbitals at a given
site. There exist other nonbonding oxygen bands and especially that one with
lowest binding energy at $(\pi,\pi)$ (seen experimentally in Ref.\
\onlinecite{pothuizen}) is not present in our model.

Now, we discuss the lowest spin polaron band $\Omega_{{\bf k}}$. Comparison
of Figs.\ 1(a) and 1(b) shows that the direct oxygen-oxygen hopping is
responsible for the dispersion along the line $\left( 0,\pi \right) -\left(
\pi ,0\right) $. The dispersion dependence on frustration in spin subsystem
and on temperature was extensively studied in our previous works \cite
{bkm2,barVan} within an extended basis set. From Fig.\ 2 we see that the
pole strength $Z\left( {\bf k}\right) $ vanishes at the $\Gamma $ point
(with ${\bf k}=\left( 0,0\right) $). This result is the consequence of a
vanishing hybridization between oxygen and copper states at this point.
Furthermore, the triplet and the nonbonding oxygen bands are degenerate
there. Consequently, all the orbitals which are incorporated into our model
give rise to only one oxygen derived peak in the photoemission spectra at the
$\Gamma$
point. And indeed, the spin polaron is not visible in the photoemission
spectra there. \cite{wells,pothuizen} Note that the vanishing of the
hybridization, and consequently the vanishing of the spin polaron is ignored
in the course of Emery model reduction to the one-band Hamiltonian (see
Refs.\ \onlinecite{yushankhai,ZhR} and references therein), and thus the
hole spectral weight is nonzero at ${\bf k}=\left( 0,0\right) $ in the $t$-$J
$ and other one-band models. Even the theories for the three-band Emery
model which start from the N\'eel order \cite{kabanov,starykh} cannot obtain
zero spectral weight at ${\bf k}=(0,0)$ since they always deal with a linear
combination of states with momentum ${\bf k}$ and ${\bf k} + {\bf q}_0$ ($
{\bf q}_0= \left( \pi,\pi \right)$.

We note also an apparent similarity of the polaron branch $\Omega _{{\bf k} }
$ of the present spectrum with the calculations based on the N\'{e}el state
and linear spin wave theory for $t$-$J$ \cite
{schmitt,kane,eder,martinez,plakida} and three-band \cite
{eder2,kabanov,starykh} models. However, our results show a slight deviation
from the symmetry of the magnetic Brillouin zone $\Omega _{{\bf k}} \neq
\Omega _{{\bf k+q}_0}$ due to the spin singlet state of the magnetic
background. We see from Fig.\ 1 that the mean field polaron bandwidth $w\sim
\tau $ ( $w=2.5\tau \left( 2.4\tau \right) $ for $h=0\left( 0.3\tau \right) $
). That is larger than expected due to the following reason: in the limit of
small $J/\tau \ll 1$ a state $\left| {\cal B}_{{\bf k}}\right\rangle ={\cal B
}_{{\bf k}}^{\dagger }| {\cal G} \rangle $ for a momentum ${\bf k}$ near $
(0,0)$ or $(\pi ,\pi )$ far from the band bottom has an energy $\Omega
_{\max }\sim \Omega _{\min }+w$ . This energy may be much greater than that
of the states
$$
|{\cal Y}_{{\bf k,q}}\rangle = \tilde S_{{\bf q}} {\cal B}_{{\bf k-q}
}^{\dagger }| {\cal G} \rangle
$$
where we defined
\begin{equation}
\tilde S_{{\bf q}}=\frac 1{\sqrt{N}}\sum_{{\bf R}}{\rm e}^{\imath {\bf qR}}
\tilde S_{{\bf R} }.
\end{equation}
The energy of the states $|{\cal Y}_{{\bf k,q}}\rangle$ is of the order of $
\Omega _{\min }+J$ for ${\bf k-q}$ near the band bottom. It means that $
\left| {\cal B}_{{\bf k}}\right\rangle $ is unstable with respect to the
decay into $|{\cal Y}_{{\bf k,q}}\rangle $ states and analogous states,
which contain more spinwaves. These states contain spin distortions that are
situated far from the hole. In order to take them into account within the PM
we should extend the basis set up to infinity. On the other hand the effect
of these states can be described by treating the scattering of the spin
polaron in terms of irreducible GF, also known as the Tserkovnikov
technique. As we will see, this repairs the too large polaron bandwidth in
the simple projection method.

\section{Scattering for polaron Green's function}
\pppp

We treat the polaron operator ${\cal B}_{{\bf k}}^{\dagger }={\cal B}_{1,
{\bf k}}^{\dagger }$ (\ref{eio}) as a candidate for the elementary
excitation and calculate the corresponding two-time retarded GF $G_p({\bf k}
,\omega )$ which is defined in (\ref{pGf}). The Dyson equation for $G_p$ has
the form
\begin{equation}
\label{dy}G_p^{-1}({\bf k},\omega )=\left[ G_p^{(0)}\right] ^{-1}-\Sigma (
{\bf k},\omega );\quad \Sigma ({\bf k},\omega )=\langle {\cal R}_{{\bf k}}|
{\cal R}_{{\bf k}}^{\dagger }\rangle _\omega ^{(irr)},
\end{equation}
with
\begin{equation}
G_p^{(0)}=(\omega -\Omega _{{\bf k}})^{-1};\quad {\cal R}_{{\bf k}}=[{\cal B}
_{{\bf k}},H]
\end{equation}
and where we used the irreducible GF
\begin{equation}
\label{Rirr}\langle {\cal R}_{{\bf k}}|{\cal R}_{{\bf k}}^{\dagger }\rangle
_\omega ^{(irr)}=\langle {\cal R}_{{\bf k}}|{\cal R}_{{\bf k}}^{\dagger
}\rangle _\omega -\langle {\cal R}_{{\bf k}}|{\cal B}_{{\bf k}}^{\dagger
}\rangle _\omega \frac 1{\langle {\cal B}_{{\bf k}}|{\cal B}_{{\bf k}
}^{\dagger }\rangle _\omega }\langle {\cal B}_{{\bf k}}|{\cal R}_{{\bf k}
}^{\dagger }\rangle _\omega
\end{equation}
in accordance with the definition given by Tserkovnikov. \cite{tserkovnikov}
The results for $G_p({\bf k},\omega )$ can be used to calculate $G_h({\bf k}
,\omega )$ for low energies
\begin{equation}
\label{ghgp}G_h({\bf k},\omega )\approx Z\left( {\bf k}\right) G_p\left(
{\bf k},\omega \right) \;.
\end{equation}
It must be underlined that Eq.\ (\ref{dy}) coincides only formally with the
Dyson equation for the causal Green's function. The ''self-energy'' $\Sigma (
{\bf k},\omega )$ in (\ref{dy}) has no diagrammatic representation and it is
represented by more complex Green's functions in the right hand side of Eq.\
(\ref{Rirr}). The derivation of Eq.\ (\ref{Rirr}) and its relation to the
conventional projection technique \cite{mori,FuldeB,forster} are represented
in Appendix A. As follows from (\ref{dy}) ${\rm Re}\ \Sigma $ gives the
renormalization of the energy of the polaron ${\cal B}_{{\bf k}}^{\dagger }$
and ${\rm Im}\ \Sigma $ gives its damping when $\left| {\rm Im}\ \Sigma (
{\bf k},\omega )\right| \ll \left| {\rm Re}\ \Sigma ({\bf k},\omega )\right|
$. In a general case the elementary excitations must be investigated
self-consistently. The self-energy $\Sigma ({\bf k},\omega )$ is dictated by
the interaction of the polaron with the spin subsystem, i.e., by the polaron
scattering on the spin waves. For this reason the main problem of the
present technique consists in calculating the irreducible Green's function (
\ref{Rirr}).

\section{Vertex calculation}
\pppp

To determine ${\cal R}_{{\bf k}}$ we have to calculate the commutator with
the Hamiltonian. That gives in detail
\begin{eqnarray}
\left[ B_{1,{\bf k}},\hat H\right] &=&\left( 2\tau \beta _{{\bf k}}^2-\frac{
8h\pi _{{\bf k}}^2}{\beta _{{\bf k}}^2}\right) B_{1,{\bf k}}+4\tau \beta _{
{\bf k}}\nu _{{\bf k}}B_{2,{\bf k}}-\frac{4h\pi _{{\bf k}}\delta _{{\bf k}}}{
\beta _{{\bf k}}^2}B_{3,{\bf k}} \; ,
\nonumber \\
\left[ B_{2,{\bf k}},\hat \tau \right]  \  &=&\tau \left[ 3\frac{\beta _{{\bf
k}}}{ \nu _{{\bf k}}}B_{1,{\bf k}}-2B_{2,{\bf k}}+\frac 2{\nu _{{\bf
k}}\sqrt{N}}
\sum_{{\bf q}}\gamma _{{\bf k-q}}\left( \beta _{{\bf k-q}}Y_{1,{\bf k,q}
}+2\nu _{{\bf k-q}}Y_{2,{\bf k,q}}\right) \right] \; ,
\nonumber \\
\left[ B_{2,{\bf k}},\hat J\right] &=& 4J\frac 1{\nu _{{\bf k}}\sqrt{N}}\sum_{
{\bf q}}\gamma _{{\bf q}}\sqrt{\frac 23}\nu _{{\bf k-q}}Y_{J,{\bf k,q}} \; ,
\nonumber \\
\left[ B_{2,{\bf k}},\hat h\right] &=&-2hB_{2,{\bf k}}+\frac 1{\nu _{{\bf k}}
\sqrt{N}}\sum_{{\bf q}}\left[ -\frac{\left( 8h\pi _{{\bf k-q}}^2-2h\beta _{
{\bf k-q}}^2\right) }{\beta _{{\bf k-q}}}Y_{1,{\bf k,q}}-\frac{4h\pi _{{\bf k-q
}}\delta _{{\bf k-q}}}{\beta _{{\bf k-q}}}Y_{3,{\bf k,q}}\right] \; ,
\label{vc}
\end{eqnarray}
where we introduced the notation
\begin{equation}
Y_{i,{\bf k,q}}=\tilde S_{{\bf q}}B_{i,{\bf k-q}}\;.
\end{equation}
In the scattering term of the exchange energy $\hat J$ arises
\begin{equation}
Y_{J,{\bf k,q}}\equiv S_{{\bf q}}^\alpha \left[ \frac 1{\nu _{{\bf k-q}}
\sqrt{\frac 23N}}\sum_{{\bf R}}{\rm e}^{-\imath \left( {\bf k-q}\right) {\bf
R}}\left( \imath \epsilon _{\alpha \beta \gamma }S_{{\bf R}}^\beta \hat
\sigma ^\gamma p_{{\bf R}}\right) \right] .
\end{equation}
which has a different structure ($\epsilon _{\alpha \beta \gamma }$ is the
antisymmetric tensor). Therefore, it has to be projected
\begin{equation}
\label{proJ}\left[ B_{2,{\bf k}},\hat J\right] \approx 4J\frac 1{\nu _{{\bf k
}}\sqrt{N}}\sum_{{\bf q}}\gamma _{{\bf q}}\left( \frac 23\right) \nu _{{\bf
k-q}}Y_{2,{\bf k,q}},
\end{equation}
using
\begin{equation}
\left\langle \left\{ Y_{J,{\bf k,q}},Y_{2,{\bf k,q}}^{\dagger }\right\}
\right\rangle =\frac 23C_{{\bf q}};\quad C_{{\bf q}}=\sum_{{\bf R}}{\rm e}
^{-\imath {\bf q}\left( {\bf n-R}\right) }\left\langle S_{{\bf n}}^\alpha S_{
{\bf R}}^\alpha \right\rangle .
\end{equation}
Finally one finds
\begin{equation}
\label{b3c}\left[ B_{3,{\bf k}},\hat H\right] =-\frac{4h\pi _{{\bf k}}\delta
_{{\bf k}}}{\beta _{{\bf k}}^2}B_{1,{\bf k}}+\frac{8h\pi _{{\bf k}}^2}{\beta
_{{\bf k}}^2}B_{3,{\bf k}}\;.
\end{equation}
After carrying out the projection (\ref{proJ}), all terms $R_{i,{\bf k}
}\equiv \left[ B_{i,{\bf k}},\hat H\right] $ can be represented in unique
form
\begin{equation}
\label{Rifull}R_{i,{\bf k}}=\lambda _{ij}B_{j,{\bf k}}+\frac 1{\sqrt{N}}
\sum_{{\bf q}}g_{ij,{\bf k,q}}Y_{j,{\bf k,q}}
\end{equation}
with coefficients $\lambda _{ij}$ and $g_{ij,{\bf k,q}}$ that can be simply
derived from (\ref{vc},\ref{proJ}) and (\ref{b3c}). One should note that the
scattering has its origin mainly in the second basis operator $B_{2,{\bf k}}$
(\ref{b1}) since only coefficients $g_{2j,{\bf k,q}}$ are different from
zero. We find from Eq.\ (\ref{Rifull}):
\begin{equation}
\label{rby}{\cal R}_{{\bf k}}=\alpha _i^{(1)}({\bf k})R_{i,{\bf k}}=\alpha
_i^{(1)}({\bf k})\lambda _{ij}B_{j,{\bf k}}+\check R_{{\bf k}}\;,
\end{equation}
where we define
\begin{equation}
\label{rsb}\check R_{{\bf k}}=\frac 1{\sqrt{N}}\sum_{{\bf q}}{\alpha
_i^{\left( 1\right) }\left( {\bf k}\right) }g_{ij,{\bf k,q}}Y_{j,{\bf k,q}}=
\frac 1{\sqrt{N}}\sum_{{\bf q}}\tilde S_{{\bf q}}\left( \alpha _i^{\left(
1\right) }\left( {\bf k}\right) g_{ij,{\bf k,q}}B_{j,{\bf k-q}}\right) \;.
\end{equation}
We see from Eqs.\ (\ref{dy},\ref{Rirr}), that the self-energy $\Sigma ({\bf k
},\omega )$ accounting for interaction effects is expressed through higher
order Green's functions (see Appendix A). One should notice that the terms
linear in ${\cal B}_{{\bf k}}\equiv {\cal B}_{1,{\bf k}}$ in Eq.\ (\ref{rby}
) do not contribute to the irreducible Green's function (\ref{Rirr}) for $
\Sigma \left( {\bf k},\omega \right) $. The terms $\propto {\cal B}_{j,{\bf k
}}$, j= 2, 3, in Eq. (\ref{rby}) are orthogonal to ${\cal B}_{{\bf k}}$ and
give nonvanishing contribution to the irreducible Green's function (\ref
{Rirr}) only in the energy region of the upper polaron bands. If the lowest
polaron band may be regarded as isolated (i.e.\ if it is energetically well
separated from other bands), we can neglect the interband scattering and
retain only the intraband one. This is the case in our problem, since the
energy gap ($1.3\tau $ for $h=0$ and $1.9\tau $ for $h=0.3$, see Fig.1)
between the two lowest bands is about half of the mean field bandwidth of
the polaron. Then we project the operator $\left( \alpha _i^{\left( 1\right)
}\left( {\bf k}\right) g_{ij,{\bf k,q}}B_{j,{\bf k-q}}\right) $ onto ${\cal B
}_{{\bf k-q}}$ and obtain finally
\begin{equation}
\label{Rw}\check R_{{\bf k}}\approx \frac 1{\sqrt{N}}\sum_{{\bf q}}\Gamma (
{\bf k,q})\tilde S_{{\bf q}}{\cal B}_{{\bf k-q}}
\end{equation}
where
\begin{equation}
\Gamma ({\bf k,q})=\left\langle \left\{ \alpha _i^{\left( 1\right) }\left(
{\bf k}\right) g_{ij,{\bf k,q}}B_{j,{\bf k-q}},\alpha _l^{\left( 1\right)
}\left( {\bf k-q}\right) B_{l,{\bf k-q}}^{\dagger }\right\} \right\rangle
=\alpha _i^{\left( 1\right) }\left( {\bf k}\right) g_{ij,{\bf k,q}}\alpha
_j^{\left( 1\right) *}\left( {\bf k-q}\right) \;.
\end{equation}
In the following we will use $\check R_{{\bf k}}$ instead of ${\cal R}_{{\bf
k}}$ in Eq.\ (\ref{Rirr}). Then the lowest order self-energy contribution is
provided by the first term in the right hand side of the expression (\ref
{Rirr}), which is responsible for the scattering of polarons, while the
second term leads to higher order corrections (see the discussion in the
Appendix A). One should mention that the operators $Y_{i,{\bf k,q}}$ are
orthogonal among each other. However, unlike the basis operators (\ref{b1},
\ref{b3}) only up to terms of order $1/N$. Therefore, one may obtain
artificial terms in the vertex $\Gamma ({\bf k,q})$ if one carries out the
calculation in a changed order than that one presented here.

In our previous work \cite{barabanov97,barjetp97} we used only one operator
in the PM. The present method generalizes the recipe of vertex calculation
from Ref.\ \onlinecite{barabanov97} to the more complex case of several
operators in the PM basis set.

\section{Self-consistent Born approximation}
\pppp

The self-energy (\ref{Rirr}) with $\check R_{{\bf k}}$ (\ref{Rw}) provides
the opportunity to apply the mode-coupling approximation in terms of an
independent propagation of the polaron and spin excitations. It consists in
the proper decoupling procedure for the two time correlation function
\begin{equation}
\label{decup}\langle \check R_{{\bf k},\sigma }(t)\check R_{{\bf k},\sigma
}^{\dagger }(t^{\prime })\rangle \simeq \frac 1N\sum_{{\bf q}}\Gamma ^2({\bf
k},{\bf q})\langle {\cal B}_{{\bf k}-{\bf q},\sigma }(t){\cal B}_{{\bf k}-
{\bf q},\sigma }^{\dagger }(t^{\prime })\rangle \langle {\bf S}_{-{\bf q}}(t)
{\bf S}_{{\bf q}}(t^{\prime })\rangle ,
\end{equation}
\begin{equation}
\label{SSq}\langle {\bf S}_{-{\bf q}}(t){\bf S}_{{\bf q}}(t^{\prime
})\rangle =\frac 1N\sum_{{\bf r},{\bf r}^{\prime }}{\rm e}^{i{\bf q\cdot }(
{\bf r}^{\prime }-{\bf r})}\left\langle {\bf S}_{{\bf r}}(t){\bf S}_{{\bf r}
^{\prime }}(t^{\prime })\right\rangle .
\end{equation}
In the frameworks of such a mode-coupling approximation the second term in
the right hand side of Eq.\ (\ref{Rirr}) may be neglected (see the
discussion in the Appendix A). The first term $\langle \check R_{{\bf k}}|
\check R_{{\bf k}}^{+}\rangle _\omega $ of (\ref{Rirr}) may be expressed in
terms of the Fourier transform of the above two-time correlation function (
\ref{decup}). The presence of the polaron-polaron correlation function $
\langle {\cal B}_{{\bf k}-{\bf q}}(t){\cal B}_{{\bf k}-{\bf q}}^{\dagger
}(t^{\prime })\rangle $ in the right hand side of Eq.\ (\ref{decup}) allows
the self-consistent calculation of $G_p({\bf k},\omega )$. Note that in the
case of the $t$-$J$ model, investigated in terms of spinless holes, the
analogous decoupling procedure for the irreducible Green's function of the
form (\ref{Rirr}) is equivalent to SCBA in a usual diagrammatic technique.
\cite{plakida}

The spin-spin correlation function $\langle {\bf S}_{-{\bf q}}(t){\bf S}_{
{\bf q}}(t^{\prime })\rangle $ in (\ref{decup}) is calculated from the spin
excitation Green's function $D({\bf q},\omega )$. We treat the spin
subsystem in the spherically symmetric approach, \cite
{kondo,shimahara,barabanov94} which is briefly described in the Appendix B.
The Green's function $D({\bf q},\omega )$ has the form
\begin{equation}
\label{gfspin}D({\bf q},\omega )=\langle S_{-{\bf q}}^\alpha |S_{{\bf q}
}^\alpha \rangle _\omega =-8JC_{{\bf g}}\frac{1-\gamma _{{\bf q}}}{\omega
^2-\omega _{{\bf q}}^2};
\end{equation}
$$
\omega _{{\bf q}}^2=-32J\alpha _1\left( C_{{\bf g}}/3\right) (1-\gamma _{
{\bf q}})(2\Delta +1+\gamma _{{\bf q}}).
$$
We neglect the influence of doped holes on the copper spin dynamics and take
the spin spectrum parameters calculated in Ref.\ \onlinecite{barabanov94}: $
\Delta =0,$ for $T=0 $, the vertex correction $\alpha_1=2.35 $, the spin
excitations condensation part $m^2=0.09$. Note that the spin excitation
spectrum $\omega _{{\bf q}}$ has the magnetic BZ symmetry at $\Delta =0,$
but the Green's function $D({\bf q},\omega )$ has the symmetry of the full
BZ due to the numerator $({1-\gamma _{{\bf q}}})$.

As a result of the decoupling (\ref{decup}) we come to the integral equation
for the Green's function
\begin{equation}
\label{gineq}G_p({\bf k},\omega )=\frac 1{\omega -\Omega _{{\bf k}}-\Sigma (
{\bf k},\omega )},
\end{equation}
where
\begin{equation}
\label{gineq1}\Sigma ({\bf k},\omega )= \frac{1}{N} \sum_{{\bf q}}M^2({\bf k}
,{\bf q })G_p({\bf k}-{\bf q},\omega -\omega _{{\bf q}}),
\end{equation}
\begin{equation}
\label{vert}M^2({\bf k},{\bf q})=\Gamma ^2({\bf k},{\bf q})\frac{\left( -4C_{
{\bf g}}\right) \left( 1-\gamma _{{\bf q}}\right) }{\omega _{{\bf q}}}.
\end{equation}
$\Gamma ({\bf k},{\bf q})$ corresponds to the bare vertex for the coupling
between a spin polaron and a spin wave. It is known \cite
{schrieffer,chubukov97} that this vertex is substantially renormalized for $
{\bf q}$ close to the antiferromagnetic vector ${\bf q}_0=(\pi ,\pi )$ .
This renormalization is due to the strong interaction of a polaron with the
condensation part of spin excitations that must be taken into account from
the very beginning. As a result, the renormalized vertex $\tilde \Gamma (
{\bf k},{\bf q})$ must be proportional to \cite{schrieffer} $\left[ \left(
{\bf q}-{\bf q}_0\right) ^2+L_s^{-2}\right] ^{1/2}$, $L_s$ being the
spin-spin correlation length, $L_s\rightarrow \infty $ in our case of a long
range order state of the spin subsystem. Below, this renormalization is
taken into account empirically by the substitution
\begin{equation}
\label{renorm}\Gamma ({\bf k},{\bf q})\rightarrow \tilde \Gamma ({\bf k},
{\bf q})=\Gamma ({\bf k},{\bf q})\sqrt{\left( 1+\gamma _{{\bf q}}\right) .}
\end{equation}
The introduced vertex correction is proportional to $\left| {\bf q}-{\bf q}
_0\right| $ for ${\bf q}$ close to ${\bf q}_0$ . Let us mention that the
bare vertex leads to a dramatic decrease of the QP bandwidth.

\section{Results and Discussion}
\pppp

In this section we present the results for the low energy part of the
one-particle retarded Green's function $G_h\left( {\bf k},\omega \right) $ (
\ref{Ghh}). In our treatment we ignore the scattering of a polaron to the
upper bands. In this approximation the low energy part of $G_h\left( {\bf k}
,\omega \right) $ is related with the polaron Green's function $G_p\left(
{\bf k},\omega \right) $ through Eq.\ (\ref{ghgp}). We are mostly interested
in the spectral function of the polaron or hole GF, respectively, which are
given by
\begin{equation}
\label{Ak}A_{p/h}\left( {\bf k},\omega \right) =-\frac 1\pi {\rm Im}\
G_{p/h}\left( {\bf k},\omega \right) \;.
\end{equation}
The spectral function of a bare hole $A_h$ is roughly proportional to the
intensity in an angle resolved photoemission (ARPES) experiment. The
self-consistent equation (\ref{gineq}) was solved by means of the recursive
procedure that provides step by step the coefficients $a_n,b_n$ of the
continued fraction expansion of $G_p\left( {\bf k},\omega \right) $. The
details of the calculation will be published elsewhere \cite{barjetp97}. The
two-dimensional Simpson's rule was used for the integration in Eq.\ (\ref
{gineq1}) over $231$ points in the irreducible part of the Brillouin zone.
The prominent feature of the expansion is the fast convergence of
coefficients $a_n,b_n$ to the asymptotic behavior which is characterized by
a linear dependence on $n$. The slope for $a_n$ is twice as large as the
slope for $b_n$. This feature allows to use the incomplete gamma function as
an analytic terminator \cite{t1} for the continued fraction. We applied the
terminator after the calculation of $30$ pairs of coefficients. Our
calculations for various models of strongly correlated electrons indicate
that the linear growth of continued fraction coefficients seems to be a
common feature of such models.

The solid line in the Figure 1 shows the dispersion $\varepsilon \left( {\bf
k} \right) $ of the lowest peak in the spectral density. First, we may see
that the overall shape of the dispersion curve has not changed in comparison
with the mean field result. On the other hand, we may note that the energy
renormalization for the top of the band is larger than for the bottom
resulting in substantial narrowing of the band. Although the band width of
the polaron in mean field $\Omega_{{\bf k}}$ is proportional to the hopping
amplitude $\tau$, it is reduced to a band width proportional to $J$ due to
the scattering of the polaron at spin fluctuations which are inherent in $
\varepsilon \left( {\bf k} \right)$. Let us note that the polaron mean field
energy $\Omega_{{\bf k}}$ represents the center of gravity of the polaron
spectral function $A_p\left( {\bf k} ,\omega \right) $ that is
\begin{equation}
\label{cgr}\Omega _{{\bf k}}=\int_{-\infty }^\infty \omega A_p\left( {\bf k}
,\omega \right) d\omega .
\end{equation}
On the other hand, we see from the explicit expressions for the Liouvillean
matrix elements (\ref{Liu}) that $\Omega _{{\bf k}}$ is unambiguously
related to the nearest neighbor and next-nearest neighbor static spin-spin
correlation functions of the undoped system. The latter is described by the
Heisenberg Hamiltonian (\ref{J}) on the square lattice, and the correlation
functions are known at present time with a very high accuracy from various
analytic \cite{shimahara,fulde} and numerical \cite{liang,oitmaa}
approaches. That means the asymmetry of $\Omega_{{\bf k}}$ (with respect to
the magnetic Brillouin zone) is not only a consequence of the particular
approximations made in the present work, but already a property of the
polaron spectral function itself.

Figure 3 gives the polaron and hole spectral densities for three
characteristic momenta ${\bf k}$ to distinguish three regimes. In addition,
we show the real and imaginary parts of the self-energy $\Sigma \left( {\bf
k },\omega \right)$ which are approximately identical for polaron or hole GF
according to Eq.\ (\ref{ghgp}). Let us recall that we have made the change $
{\bf k\rightarrow k}^{\prime }={\bf k}-{\bf q}_0$ in the final results, so
that the quasimomenta in the figures correspond to the actual Brillouin zone
of the ${\rm CuO}_2$ plane, in contrast to our previous work. \cite
{barabanov97} We find a sharp quasiparticle peak at the bottom of the
spectra for ${\bf k}_a=\left( \pi /2,\pi /2\right) $ and ${\bf k}_b=\left(
7\pi /20,7\pi /20\right) $. The position of the quasiparticle peak
corresponds to the condition ${\rm Re} \ G_p^{-1}\left( {\bf k},\omega
\right) =0$, i.e. the point where we have a crossing of the functions $
\omega -\Omega _{{\bf k}}$ and ${\rm Re} \ \Sigma \left( {\bf k},\omega
\right)$, see Figs.\ 3(a) and 3(b). Quite a different shape has the spectral
density$A_p\left( {\bf k}_c,\omega \right)$ at ${\bf k}_c=\left( 0,0\right)$
. Its lowest peak is not related to the zero of ${\rm Re} \ G_p^{-1}\left(
{\bf k} ,\omega \right) $ and is due to abrupt changes in the self-energy.
That leads to a large imaginary part of $\Sigma \left( {\bf k}, \omega
\right)$ at the position of the lowest peak in the spectral function. We may
say that there is {\em no} quasiparticle excitation for this quasimomentum
value.

Figure 4 shows the difference between the quasiparticle excitations at
points ${\bf k}_a$ and ${\bf k}_b$ more in detail. The first one corresponds
to a quasiparticle with infinite lifetime, i.e. to an isolated pole of the
Green's function at the real axis of the complex energy plane. For ${\bf k}
_b $ the situation is different, a very small, but finite, imaginary part of
the self-energy is present in the region of the peak. The shape of the peak
deviates from Lorentzian. Therefore, such an excitation can be regarded as a
quasiparticle with finite lifetime.

Table 1 shows the position of the lowest peak, $\varepsilon \left( {\bf k}
\right) $, the area under the peak for polaron $W_p\left({\bf k}\right)$ and
for hole spectral function, $W_h\left( {\bf k}\right) =Z\left( {\bf k}
\right) W_p\left( {\bf k}\right) $ (see (\ref{ghgp})), and the value of the
imaginary part of the self-energy, $-{\rm Im} \ \Sigma \left[ {\bf k}
,\varepsilon \left( {\bf k}\right) \right] $ for ${\bf k}$ values which vary
along the diagonal of the Brillouin zone. The rows which are marked by a
star do not represent a quasiparticle excitation like at ${\bf k}_c$ and the
filled or open circles correspond to momenta with quasiparticles of infinite
or finite lifetime, respectively.

So, we may distinguish three qualitatively different kinds of spectral
function $A_p\left( {\bf k},\omega \right) $ behavior:

(i) The lowest peak tends to a delta function $W_p\delta \left[ \omega
-\varepsilon\left( {\bf k}\right) \right] $, when ${\rm Im} \ \omega
\rightarrow 0$ (see Fig.4(a)). We have a quasiparticle, characterized by the
excitation energy $\varepsilon\left( {\bf k} \right) =\Omega _{{\bf k}}+{\rm
Re} \ \Sigma \left[ {\bf k},\varepsilon \left( {\bf k}\right) \right] $ with
infinite lifetime $\tau _l$, $\left( 1/\tau_l\right) \equiv -{\rm Im} \
\Sigma \left[ {\bf k}, \varepsilon\left( {\bf k} \right) \right] =0$. These $
{\bf k}$-points are situated near the band minimum ${\bf k}_{\min }$. They
are marked by solid circles at Fig.\ 1 and in Table 1.

(ii) The lowest peak has approximate Lorentzian form (see Fig.4(b))
\begin{equation}
\label{lor}A_p\left( {\bf k},\omega \right) = - \frac{1}{\pi} {\rm Im} \ G_p
\left( {\bf k},\omega \right) \approx W_p \frac { \left( 1/\tau_l\right)} {
\left[ \omega - \varepsilon\left( {\bf k}\right) \right] ^2+\left( 1/\tau
_l\right) ^2} + A_{incoh}
\end{equation}
and corresponds to a quasiparticle with energy $\varepsilon \left( {\bf k}
\right) =\Omega _{{\bf k}}+{\rm Re} \ \Sigma \left[ {\bf k}, \varepsilon
\left( {\bf k}\right) \right] $, and finite lifetime $\left( 1/\tau_l
\right) \approx -{\rm Im} \ \Sigma \left[ {\bf k},\varepsilon \left( {\bf k}
\right) \right] \ll \left| \varepsilon \left( {\bf k}\right) - \varepsilon
\left( {\bf k}_{\min }\right) \right| $. The corresponding ${\bf k}$-values
are marked by open circles at Fig.\ 1 and in Table 1.

(iii) The spectral density is completely incoherent and is of the same order
of magnitude as the imaginary part of the self-energy at all $\omega $ (see
Fig.3(c)). These ${\bf k}${\bf -}points are marked by stars.

In order to understand the nature of the polaron damping ($1/\tau_l$ in (\ref
{lor})), let us consider a particular ${\bf k}$-value and estimate the
imaginary part of the self-energy at a particular $\omega $. Changing the
summation variable in (\ref{gineq1}) we can calculate the imaginary part of
the self-energy as follows
\begin{equation}
\label{damping}-{\rm Im} \ \Sigma ({\bf k},\omega )= \pi \frac{1}{N} \sum_{
{\bf q} ^{\prime }} M^2({\bf k},{\bf k}- {\bf q}^{\prime }) A_p \left( {\bf q
}^{\prime}, \omega -\omega _{{\bf k}-{\bf q}^{\prime}} \right) \; .
\end{equation}
Now we suppose that the lowest peak at all ${\bf q}^{\prime }= {\bf k-q}$
may be described approximately by Lorentzian (\ref{lor}). Then we see from
Eq.\ (\ref{damping}) that large damping, characterized by a large value of $
- {\rm Im} \ \Sigma ({\bf k},\omega )$ may only arise when the condition
\begin{equation}
\label{cdam}\omega -\varepsilon \left( {\bf q}^{\prime }\right) -\omega _{
{\bf k-q}^{\prime }}=0
\end{equation}
is satisfied for some values of ${\bf q}^{\prime}$ (see also Ref.\
\onlinecite
{yushankhai}). On the contrary, if Eq.\ (\ref{cdam}) does not hold even at
the lowest edge of the spectral function $\omega =\varepsilon \left( {\bf k}
\right) $, the damping is absent. In Fig.5 we plot $\varepsilon \left( {\bf
q }^{\prime }\right) $ together with the curve $\varepsilon \left( {\bf k}
\right) -\omega _{{\bf k-q}^{\prime }}$ for various values of ${\bf k} $
(for $h=0.3\tau $ and where we choose only one direction for both ${\bf k}$,
and ${\bf q}^{\prime}$ ). We may see that for ${\bf k}_{\min }{\bf =}(\pi
/2,\pi /2)$ the condition (\ref{cdam}) holds only for the trivial values $
{\bf q}={\bf k}-{\bf q}^{\prime}=(0,0)$ and ${\bf q=q}_0$, where, however,
the vertex $M^2({\bf k},{\bf q})$ vanishes. It demonstrates that it is
impossible to scatter from ${\bf k}_{\min }$ into other states. The same
situation occurs for some finite region of ${\bf k}$ values around the band
bottom. These are exactly those momenta with a quasiparticle of infinite
lifetime. Different is the situation for ${\bf k}{\bf =}(0,0)$ and ${\bf k=}
(4\pi /20,4\pi /20)$ where several nontrivial intersections take place.

Figure 6 shows the low energy part of the hole spectral density $A_h\left(
{\bf k},\omega \right) $ for ${\bf k}$ along high symmetry directions. Let
us recall that the intensity of the lowest peak of $A_h\left( {\bf k},\omega
\right) $ is governed by two circumstances: the ${\bf k}$-dependence of the
mean field residue $Z\left( {\bf k}\right) $ (see Eq.\ (\ref{Ghp}) and the
discussion about Fig.\ 2) and the intensity of the lowest polaron peak of $
A_p\left( {\bf k},\omega \right) $. One can clearly observe a well defined
quasiparticle peak near the bottom of the spectrum at $(\pi /2,\pi /2)$,
everywhere along the line $(0,\pi )$-$(\pi ,0)$ and also near to $(\pi ,0)$.
At the $\Gamma $-point $(0,0)$, there is no hole spectral density due to the
vanishing residue for the hole Green's function in the lowest mean field
polaron band. At the same time one observes intensity at $(\pi ,\pi )$. A
clear asymmetry of the peak intensity is seen along the diagonal of the BZ.
The abrupt drop of the peak intensity in the region $(\pi /2,\pi /2)$ -$(\pi
,\pi )$ is related to the strong polaron damping there. In other words, most
of the mean field hole spectral weight ($Z({\bf k})$ in Fig.2) goes into the
incoherent part of the spectrum (situated at higher energies) due to the
strong coupling of the polaron ${\cal B}_{{\bf k}}$ with spin excitations.
An analogous behavior of the quasiparticle peak intensity was obtained in
Ref.\ \onlinecite{eder3} within a variational ansatz for the $t$-$J$ model.

Let us compare the results with those of the usual SCBA of spinless holes in
the pure $t$-$J$ model. There, one finds quasiparticles with infinite
lifetime and finite weight everywhere in ${\bf k}$-space. Additionally, the
spectral function has the symmetry of the magnetic BZ. Introducing
additional hopping terms (corresponding to direct oxygen-oxygen hopping in
the Emery model) leads to a scattering mechanism such that the upper parts
of the spectrum loose their quasiparticle character. \cite{bala,yushankhai}
On the contrary, in our approach, the damping of the spin polaron is already
present without direct oxygen-oxygen hopping. If we consider only the
dispersion relation $\varepsilon ( {\bf k} )$ we observe a remarkable
similarity between the present calculation and earlier results using the
SCBA of spinless holes or other methods. In fact, the deviations from the
magnetic BZ symmetry in Fig.\ 1 are not very large. On the other hand, we
find some qualitatively new features of the spectral function which are
absent in the numerical results of Refs.\ \onlinecite{martinez,plakida}.
Besides the strong polaron scattering away from the band bottom we note
especially: (i) the absence of the polaron quasiparticle at the $\Gamma$
-point and (ii) the asymmetry of the peak intensity with respect to the
magnetic BZ.

Our results (Fig.\ 6) can also be summarized in such a way that a well
pronounced polaron quasiparticle peak exists only around the bottom of the
band. In the direction $(\pi/2,\pi/2)$-$(\pi,0)$ it is more clearly seen
than perpendicular to it. The recent experimental finding \cite{pothuizen}
shows, indeed, that the Zhang-Rice singlet can be observed in Sr$_2$CuO$_2$Cl
$_2$ only in a similar region of the BZ. And also the higher peak intensity
going from $(\pi/2,\pi/2)$ to the $\Gamma$ -point in comparison with the
opposite direction is in agreement with experiment. So we see that those
details in which our calculation differs from the standard one \cite
{schmitt,kane,martinez,liu,plakida} are essential for a better understanding
of the experiment.

Let us emphasize that for a detailed comparison to a real experiment there
are a considerable number of complications which have not been taken into
account in the present work: First, in general a photohole can be generated
also on copper, and there will be (in general ${\bf k}$-dependent)
interference between the photoholes created on different atoms. An operator $
c_{{\bf R+a}}^{\dagger }$ (\ref{sw}) that enters the spin-fermion
Hamiltonian (\ref{ham}) creates a hole in a state that is an antibonding (in
hole notation) combination of oxygen $p_\sigma$ state and singly occupied
copper $d_{x^2-y^2}$ state at adjacent sites. In a more realistic theory one
should deal with a complete set of bare oxygen and copper hole creation
operators. This may be expected to have some influence on the spectral
weight. Next, one must take into account interference terms like $< p_x| p_y>
$, because the photoemission operator is in general a linear combination of $
p_x$ and $p_y$. Thereby the relative phase between $p_x$ and $p_y$ depends
on the momentum transfer, and is in general even different whether one is in
the 1st or 2nd Brillouin zone etc. Clearly this will give additional
momentum dependence of the weight. Moreover, it is not even a priori clear
that $p_x$ and $p_y$ have equal weight in the PES-operator. Rather this
depends in a relatively complicated way, e.g., on the polarization of the
incoming X-ray photons. It is in fact well known experimentally that the
spectra depend quite sensitively on the X-ray polarization in several
oxychlorides like Sr$_2$CuO$_2$Cl$_2$ or Ba$_2$Cu$_3$O$_4$Cl$_2$. \cite
{schmelz,golden} At the present stage of many-body theory it is impossible
to take all these complications into account. Thus, our comparison with
experiment has only qualitative character. Nevertheless, we point out that
according to our results, many features of the experimental spectra may
originate from the peculiarities of the correlated ground state of the
material rather than from the above complications.

We may consider now what happens if we begin to dope our system. First, we
suppose that for extremely small doping the Fermi surface will consists of
hole pockets centered at $\left( \pi /2,\pi /2\right) $. The small spectral
weight of the quasiparticle poles in this region means that the surface
enclosed inside the Fermi surface may be much larger than the number of
holes. Note that due to the strong weight asymmetry the experimental
observation of the Fermi surface parts facing M point with ${\bf k}=\left(
\pi ,\pi \right) $ point may be very hard to do. In Ref.\
\onlinecite{Marshall} the observation of weak features in photoemission
spectra in this region of ${\bf k}$-space was reported, but the authors can
not unambiguously interpret them as Fermi surface cuts. With the increase of
doped holes the antiferromagnetic correlations in the spin subsystem are
weakened resulting in a deformation of the hole dispersion. \cite
{barVan,hayn} The minimum of the hole spectrum should shift to the M point
and a 'large' Fermi surface develops. An analogous scenario was also
developed in Ref.\ \onlinecite
{chubukov97}.

\section{Conclusion}

We proposed a new method to calculate the spectral function of the Emery
model in its spin-fermion form combining the Mori-Zwanzig projection method
with a representation of the self-energy by irreducible GF. Self-energy
effects reduce the mean field bandwidth of the spin polaron to a realistic
bandwidth of the order of $2J$, but do not change the overall shape of the
dispersion. Direct oxygen-oxygen hopping leads to a more isotropic band
minimum around $(\pi/2,\pi/2)$. The quasiparticle weight is maximal (about
0.25) around the minimum of the dispersion. Despite some similarities, we
observed several new features of the spectral function which are not present
in the usual SCBA \cite{schmitt,kane,martinez,liu,plakida} for spinless
holes coupled to spinons in the pure $t$-$J$ model, namely the absence of
quasiparticle weight at the $\Gamma$ point, the strong damping away from the
minimum and the asymmetry with respect to the antiferromagnetic BZ (where
the asymmetry is present, however, in other approaches to the $t$-$J$ model)
\cite{eder3,eskes}. Qualitatively, these features can be observed in the
ARPES experiments on Sr$_2$CuO$_2$Cl$_2$. We studied in detail possible
damping processes of the spin polaron and distinguished three regions in $
{\bf k}$-space where the spin polaron exists with infinite lifetime, finite
lifetime or does not exist at all.

\section*{Acknowledgements}

We are grateful to H.\ Eschrig and L.B.\ Litinski for valuable discussions
and comments. This work was supported, in part, by the INTAS-RFBR (project
No.\ 95-0591), by RSFR (Grants No.\ 98-02-17187 and 98-02-16730), by Russian
National
program on Superconductivity (Grant No. 93080), ISI Foundation and EU NTAS
Network 1010-CT930055. R.O.K. thanks the Max Planck Arbeitsgruppe
Elektronensysteme, in the Technische Universit\"at Dresden for hospitality
during accomplishing of this work.

 %%%%%%%%%%%%%%%%%%%%%%%%%%%%%%%%%%%%%%%%%%%%%%%%%%%%%%%%%%%%%%%%%%%%%%%
\newpage
\appendix
\setcounter {section} {1}

\begin{center}
{\Large {\bf Appendix A}}
\end{center}

\setcounter{pzz}{\value{equation}}\setcounter{equation}{0}
\renewcommand {\theequation}{\Alph{section}.\arabic{equation}}

 %\section{Appendix A}

Here we give the derivation of the Tserkovnikov expression for the
self-energy (\ref{Rirr}) and compare it with the conventional projection
technique expression. We use the equation of motion for the retarded Green's
function $G_p\left( {\bf k},\omega \right) =\left\langle {\cal B}_{{\bf k}
}\mid {\cal B }_{{\bf k}}^{\dagger }\right\rangle _\omega $ (\ref{pGf})
\begin{equation}
\label{A1}\omega \left\langle {\cal B}_{{\bf k}}\mid {\cal B}_{{\bf k}
}^{\dagger }\right\rangle _\omega =1+\left\langle {\cal R}_{{\bf k}}\mid
{\cal B}_{{\bf k}}^{\dagger }\right\rangle _\omega =1+\left\langle {\cal B}_{
{\bf k}}\mid {\cal R}_{{\bf k}}^{\dagger }\right\rangle _\omega .
\end{equation}
Multiplying the equation for the higher order GF
$$
\omega \left\langle {\cal R}_{{\bf k}}\mid {\cal B}_{{\bf k}}^{\dagger
}\right\rangle _\omega =\left\langle \left\{ {\cal R}_{{\bf k}},{\cal B}_{
{\bf k}}^{\dagger }\right\} \right\rangle +\left\langle {\cal R}_{{\bf k}
}\mid {\cal R}_{{\bf k}}^{\dagger }\right\rangle _\omega
$$
by $\left\langle {\cal B}_{{\bf k}}\mid {\cal B}_{{\bf k}}^{\dagger
}\right\rangle _\omega $ , and noting that $\left\langle \left\{ {\cal R}_{
{\bf k}},{\cal B}_{{\bf k}}^{\dagger }\right\} \right\rangle =\Omega _{{\bf
k }}$ , we have
\begin{equation}
\label{A2}\omega \left\langle {\cal R}_{{\bf k}}\mid {\cal B}_{{\bf k}
}^{\dagger }\right\rangle _\omega \left\langle {\cal B}_{{\bf k}}\mid {\cal
B }_{{\bf k}}^{\dagger }\right\rangle _\omega =\Omega _{{\bf k}}\left\langle
{\cal B}_{{\bf k}}\mid {\cal B}_{{\bf k}}^{\dagger }\right\rangle _\omega
+\left\langle {\cal R}_{{\bf k}}\mid {\cal R}_{{\bf k}}^{\dagger
}\right\rangle _\omega \left\langle {\cal B}_{{\bf k}}\mid {\cal B}_{{\bf k}
}^{\dagger }\right\rangle _\omega .
\end{equation}
In the left-hand side we use the second equality (\ref{A1}) to obtain
\begin{equation}
\label{A3}\left\langle {\cal R}_{{\bf k}}\mid {\cal B}_{{\bf k}}^{\dagger
}\right\rangle _\omega =\Omega _{{\bf k}}\left\langle {\cal B}_{{\bf k}}\mid
{\cal B}_{{\bf k}}^{\dagger }\right\rangle _\omega +\left\langle {\cal R}_{
{\bf k}}\mid {\cal R}_{{\bf k}}^{\dagger }\right\rangle _\omega \left\langle
{\cal B}_{{\bf k}}\mid {\cal B}_{{\bf k}}^{\dagger }\right\rangle _\omega
-\left\langle {\cal R}_{{\bf k}}\mid {\cal B}_{{\bf k}}^{\dagger
}\right\rangle _\omega \left\langle {\cal B}_{{\bf k}}\mid {\cal R}_{{\bf k}
}^{\dagger }\right\rangle _\omega ,
\end{equation}
and substitute it into first equality (\ref{A1}). Finally Eq. (\ref{A1})
takes the form
\begin{equation}
\label{A4}\left[ \omega -\Omega _{{\bf k}}-\Sigma \left( {\bf k},\omega
\right) \right] G_p\left( {\bf k},\omega \right) =1
\end{equation}
with the self energy given by (\ref{Rirr}).

In order to clarify the physical meaning of this expression let us compare
it with the close expression for $\Sigma _{PT} \left( {\bf k},\omega \right)$
which follows from the projection technique (see Refs.\
\onlinecite{mori,FuldeB,forster}). Taking the usual notations for the scalar
product in the operator space, the projection operators $P$ and $Q$ and the
Liouvillean superoperator ${\cal L}$
\begin{equation}
\left( A\mid B\right) =\left\langle \left\{ A,B\right\} \right\rangle ,\
P=\mid {\cal B}_{{\bf k}}^{\dagger })({\cal B}_{{\bf k}}\mid ,\ Q=1-P,\
{\cal L}A=\left[ H,A\right] ,\ A{\cal L}=\left[ A,H\right] ,
\end{equation}
Green's function and the self-energy may be written as
$$
G_p\left( {\bf k},\omega \right) =({\cal B}_{{\bf k}}\mid \frac 1{\omega -
{\cal L}}{\cal B}_{{\bf k}}^{\dagger })=({\cal B}_{{\bf k}}\frac 1{\omega -
{\cal L}}\mid {\cal B}_{{\bf k}}^{\dagger }).
$$
\begin{equation}
\label{sigPT}\Sigma _{PT}\left( {\bf k} ,\omega \right) =({\cal B}_{{\bf k}
}\mid {\cal L}Q\frac 1{\omega -{\cal L}Q} {\cal LB}_{{\bf k}}^{\dagger })=(
{\cal B}_{{\bf k}}{\cal L}Q\mid \frac 1{ \omega -Q{\cal L}Q}Q{\cal LB}_{{\bf
k}}^{\dagger }),
\end{equation}
The Tserkovnikov expression (\ref{Rirr}) has the following form in the
projection technique notation
$$
\Sigma _T\left( {\bf k},\omega \right) =({\cal B}_{{\bf k}}{\cal L}\mid
\frac 1{\omega -{\cal L}}{\cal LB}_{{\bf k}}^{\dagger })-({\cal B}_{{\bf k}}
{\cal L}\mid \frac 1{\omega -{\cal L}}{\cal B}_{{\bf k}}^{\dagger })\frac 1{
( {\cal B}_{{\bf k}}\mid \frac 1{\omega -{\cal L}}{\cal B}_{{\bf k}
}^{\dagger })}({\cal B}_{{\bf k}}\mid \frac 1{\omega -{\cal L}}{\cal LB}_{
{\bf k} }^{\dagger })=
$$
\begin{equation}
\label{sigTs}=({\cal B}_{{\bf k}}{\cal L}Q\mid \frac 1{\omega -{\cal L}}Q
{\cal LB}_{{\bf k}}^{\dagger })-({\cal B}_{{\bf k}}{\cal L}Q\mid \frac 1{
\omega -{\cal L}}{\cal B}_{{\bf k}}^{\dagger })\frac 1{({\cal B}_{{\bf k}
}\mid \frac 1{\omega -{\cal L}}{\cal B}_{{\bf k}}^{\dagger })}({\cal B}_{
{\bf k}}\mid \frac 1{\omega -{\cal L}}Q{\cal LB}_{{\bf k}}^{\dagger }).
\end{equation}
In the last equality we excluded the terms linear in ${\cal B}_{{\bf k}}$.
Now we see, that the first term in $\Sigma _T\left( {\bf k},\omega \right) $
describes the propagation of a ''fluctuation'' ${\cal B}_{{\bf k}}{\cal L}Q$
in the full Liouvillean space, and the second counter-term eliminates items
corresponding to the propagation in the subspace spanned by ${\cal B}_{{\bf k
}}$, i.e., it makes the same job as the projectors $Q$ surrounding ${\cal L}$
in the denominator of $\Sigma _{PT}\left( {\bf k},\omega \right) $ (\ref
{sigPT}). Using the known rules of matrix algebra it is in fact easy to see
that $\Sigma _{PT}\left( {\bf k},\omega \right)$ (\ref{sigPT}) is equivalent
to $\Sigma _{T}\left( {\bf k},\omega \right)$ (\ref{sigTs}). When we
consider only intraband scattering we approximate ${\cal B}_{{\bf k}}{\cal L}
Q\approx \check R_{{\bf k}}, $ with $\check R_{{\bf k}}$, given by Eq.\ (\ref
{Rw}). Then our set of approximations is equivalent to the mode-coupling
approximation in the projection technique. \cite{aksen}

\appendix
\setcounter {section} {2}

\begin{center}
{\Large {\bf Appendix B}}
\end{center}

\setcounter{pzz}{\value{equation}}\setcounter{equation}{0}
\renewcommand {\theequation}{\Alph{section}.\arabic{equation}}

Now we review briefly the Kondo-Yamaji theory \cite{kondo} for an isotropic
Heisenberg spin-half antiferromagnet on a square lattice. \cite
{shimahara,barabanov94}

In the absence of holes our Hamiltonian reduces to the spin Hamiltonian $
\hat J$ (\ref{J}). We search the retarded spin-spin Green's function $
D\left( {\bf q},\omega \right) \equiv \left\langle S_{{\bf q}}^\alpha \mid
S_{-{\bf q}}^\alpha \right\rangle _ \omega $ taking into account that $
\left\langle S_{{\bf R}}^\alpha \right\rangle \equiv 0 $ and $\left\langle
S_{{\bf R}}^\alpha S_{{\bf R+r} }^\beta \right\rangle ={\frac 13}\ \delta
^{\alpha \beta } C_{{\bf r}} $ in the adopted spherically symmetric
approach. Then we have the following equations of motion for the spin
Green's functions
\begin{equation}
\label{aeqm1}\omega \left\langle S_{{\bf q}}^\alpha \mid S_{-{\bf q}}^\alpha
\right\rangle _\omega =\frac 1N\sum_{{\bf n,m}}{\rm e}^{\imath {\bf q}\left(
{\bf n-m}\right) }\left[ J\imath \epsilon _{\alpha \beta \gamma }\sum_{{\bf
g }}\left\langle S_{{\bf n+g}}^\beta S_{{\bf n}}^\gamma \mid S_{{\bf m}
}^\alpha \right\rangle _\omega \right]
\end{equation}
$$
\omega \frac 1N\sum_{{\bf n,m}}{\rm e}^{\imath {\bf q}\left( {\bf n-m}
\right) }J\imath \epsilon _{\alpha \beta \gamma }\sum_{{\bf g} }\left\langle
S_{{\bf n+g}}^\beta S_{{\bf n}}^\gamma \mid S_{{\bf m} }^\alpha
\right\rangle _\omega =zJ\left| C_{{\bf g}}\right| \left( 1-\gamma_{{\bf q}
}\right) + {\frac 12}z J^2 \left( 1-\gamma_{{\bf q}}\right) \omega
\left\langle S_{{\bf q}}^\alpha \mid S_{-{\bf q}}^\alpha \right\rangle
_\omega+
$$
$$
J^2\frac 1N\sum_{{\bf n,m}}{\rm e}^{\imath {\bf q}\left( {\bf n-m}\right) }
\sum_{{\bf g}_1 \ne {\bf g}}^{ \beta \ne \alpha } \left\langle S_{{\bf n}
}^{\beta}S_{{\bf n+g}}^{\beta}S_{{\bf n+g}-{\bf g}_1}^{\alpha}+ S_{{\bf n+g}
}^{\beta}S_{{\bf n+g}_1}^{\beta}S_{{\bf n}}^{\alpha} - (S_{{\bf n}
}^{\beta}S_{{\bf n+g}_1}^{\beta}+ S_{{\bf n+g}-{\bf g}_1}^{\beta}S_{{\bf n}
}^{\beta})S_{{\bf n+g}}^{\alpha} \mid S_{{\bf m}}^\alpha \right\rangle
_\omega
$$
\begin{equation}
\label{aeqm2}\approx zJ\left| C_{{\bf g}}\right| \left( 1-\gamma _{{\bf q}
}\right) +\left[ J^2z^2\frac 23\left| C_{{\bf g}}\right| \alpha _1\left(
1-\gamma _q\right) \left( 2\Delta +1+\gamma _q\right) \right] \left\langle
S_{{\bf q} }^\alpha \mid S_{-{\bf q}}^\alpha \right\rangle _\omega
\end{equation}
$z=4$ is the coordination number. The last approximate equality was obtained
by the unique decoupling procedure for three-site (on different sites)
spin-operators:
\begin{equation}
\label{sdec}S_{{\bf n}_1}^{\beta}S_{{\bf m}}^{\beta} S_{{\bf l}}^\alpha
\approx {\frac 13}{\alpha _{1(2)}} C_{{\bf n-m}}S_{{\bf l}}^\alpha,\quad
\beta \ne \alpha.
\end{equation}
Here the parameters $\alpha _1$ or $\alpha _2$ are used for the first  and
the second nearest neighbors ${\bf n}$ and ${\bf m}$ accordingly. They are
regarded as vertex corrections which have to be introduced in order to
prevent the violation of the sum rule of the correlation function. The
energy gap parameter equals
\begin{equation}
\label{delt}2\Delta =\frac{\alpha _2\sum_{{\bf g,g}_1}C_{{\bf g+g}_1}}{
\alpha _1z^2\left| C_{{\bf g}}\right| }-\frac{\alpha _2-1}{\alpha _1}\frac 1{
2z\left| C_{{\bf g}}\right| }-\frac{z-1}z \; .
\end{equation}
The substitution of Eq.\ (\ref{aeqm2}) into Eq.\ (\ref{aeqm1}) provides $
D\left( {\bf q},\omega \right) $ in the form (\ref{gfspin}). The static
correlation functions $C_{{\bf r}}=\left\langle {\bf S}_{{\bf R}}{\bf S}_{
{\bf R+r}} \right\rangle $ may be expressed through $D\left( {\bf q},\omega
\right) $:
\begin{equation}
\label{Cr}C_{{\bf r}}=\frac 1N\sum_{{\bf q}}{\rm e}^{-\imath {\bf qr}
}\int_{-\infty }^\infty d\omega \frac{\left[ -{\rm Im} \ D\left( {\bf q}
,\omega \right) /\pi \right] }{\exp \left( \omega /T\right) -1}=\frac 1N
\sum_{{\bf q} }{\rm e}^{-\imath {\bf qr}}\left[ -zJC_{{\bf g}}\left(
1-\gamma _{{\bf q} }\right) \frac{\coth \left( \omega _{{\bf q}}/2T\right) }{
\omega _{{\bf q}}} \right]
\end{equation}
For ${\bf r}={\bf 0},\ C_{{\bf 0}}=3/4$, and the equation (\ref{Cr}) gives
the sum rule for $D\left( {\bf q},\omega \right) $. Thus we obtain the set
of self-consistent equations (\ref{gfspin}),(\ref{Cr}). Detailed
considerations show that it is overdetermined \cite{shimahara} and one extra
condition may be imposed to solve it. This condition is the relation between
$\alpha _1$ and $\alpha _2$ given below. In a macroscopic system ($
N\rightarrow \infty $) long-range-order is present for $T=0$. We have $
\Delta =0$,
$$
\omega _{{\bf q}}=zJ\left( 2\alpha _1\left| C_{{\bf g}}\right| /3\right)
^{1/2}\sqrt{1-\gamma _{{\bf q}}^2},\quad m^2=\lim _{{\bf r}\rightarrow
\infty }{\rm e}^{\imath {\bf q}_0{\bf r}}C_{{\bf r}}\neq 0.
$$
The set (\ref{Cr}) reduces to
\begin{equation}
\label{Cr0}C_{{\bf r}}=\frac 1N\sum_{{\bf q\neq q}_0}{\rm e}^{-\imath {\bf
qr }}\left[ -\frac{zJC_{{\bf g}}\left( 1-\gamma _{{\bf q}}\right) }{\omega _{
{\bf q}}}\right] +{\rm e}^{-\imath {\bf q}_0{\bf r}}m^2.
\end{equation}

Shimahara and Takada \cite{shimahara} explored various kinds of extra
conditions to determine the parameter's values and found $m$ to be the most
sensitive quantity. They proposed to fix $r_\alpha =\left( \alpha
_1-1\right) /\left( \alpha _2-1\right) =0.82579$ for all temperatures which
gives $m=0.3$ for $T=0$. Then the values of ground state energy and uniform
spin susceptibility agree with other theories and Monte-Carlo simulations
within few percents. For finite temperature the KY theory leads to the
absence of LRO and provides a spin susceptibility which has a satisfactory
behavior over the whole temperature region: it has a zero temperature
derivative at $T=0$, it exhibits a peak around $T\sim J$ and reproduces the
high temperature expansion for $T\geq 1.6J$. In the absence of LRO, the low
lying excitations are essentially spin waves propagating in a short-range
order with a temperature dependent correlation length. On the other hand, if
there exists LRO, the usual spin wave excitations are recovered.

%%%%%%%%%%%%%%%%%%%%%%%%%%%%%%%%%%%%%%%%%%%%%%%%%%%%%%%%%%%%%%%%%%%%%%%%

\begin{figure}
\caption{
The dispersion of the lowest QP band $\varepsilon \left( {\bf k} \right)$ and
the mean field
dispersions  $\Omega _{{\bf k}}^{(i)}$ along high symmetry lines in the
Brillouin zone for $J=0.2$ and $\tau=1$ without (a, $h=0$) and with (b,
$h=0.3$) direct oxygen-oxygen hopping. Filled and open circles mark
quasiparticles with infinite and finite lifetime, respectively. A star
corresponds to  a lowest
peak in the spectral density which is strongly overdamped. The points in ${\bf
k}$ space mean: $\Gamma = (0,0)$, $X=(\pi,0)$, $M=(\pi,\pi)$ and
$X^{\prime}=(0,\pi)$. }

\label{f1}

\end{figure}

\begin{figure}
\caption{
The QP weights $Z^{(i)} \left( {\bf k} \right) $ of all three bands in the
projection method (PM) and
the area under the lowest peak of the hole spectral function $W_h \left( {\bf
k} \right) $ calculating the irreducible GF of the self-energy in
self-consistent Born approximation (SCBA) along
high symmetry lines in the
Brillouin zone for $J=0.2$ and $\tau=1$ without (a, $h=0$) and with (b,
$h=0.3$) direct oxygen-oxygen hopping. }

\label{f2}

\end{figure}

\begin{figure}

\caption{
Polaron and hole spectral densities $A_{p/h} \left( {\bf k}, \omega \right) $
around the lowest peak for the momenta: a) ${\bf k}=(\pi/2,\pi/2)$, b) ${\bf
k}=7(\pi/20,\pi/20)$ and c) ${\bf k}=(0,0)$. Also shown are the imaginary part
of the self-energy and the crossing of the curves $\omega-\Omega_{{\bf k}}$
and ${\rm Re} \ \Sigma \left( {\bf k}, \omega \right) $. The spectral functions
are broadened by a small imaginary part $ \eta=0.005 $ in $ \omega $. }

\label{f3}

\end{figure}

\begin{figure}

\caption{
Polaron spectral density and self-energy for: a) ${\bf k}=(\pi/2,\pi/2)$ and b)
${\bf k}=7(\pi/20,\pi/20)$ to distinguish a quasiparticle with infinite
lifetime (a) and a finite one (b). }

\label{f4}

\end{figure}

\begin{figure}

\caption{
Crossing of the curves $\varepsilon ( {\bf q}^{\prime} )$ and $\varepsilon (
{\bf k} )
- \omega_{{\bf k}-{\bf q}^{\prime}}$ for three different values of ${\bf k}$
along the
diagonal of the BZ for ${\bf q}^{\prime}=(q_x^{\prime},q_x^{\prime})$: there
are only trivial crossing points
(${\bf q}={\bf k}-{\bf q}^{\prime}$ equal to $\Gamma$ or M) for ${\bf
k}=(\pi/2,\pi/2)$ but nontrivial crossings for ${\bf
k}=(4\pi/20,4\pi/20)$ and ${\bf k}=(0,0)$.  }

\label{f5}

\end{figure}

\begin{figure}

\caption{
Contour plot of hole spectral function $A_{h} \left( {\bf k}, \omega \right) $
for for $J=0.2$, $\tau=1$,  $h=0.3$ and $\eta=0.01$ along high symmetry lines
in the BZ a) $(0,0)$-$(\pi,\pi)$, b) $(\pi,0)$-$(0,\pi)$ and c)
$(0,0)$-$(\pi,0)$. }

\label{f6}

\end{figure}

\newpage

\begin{table}

\caption{Position of the lowest energy peak  $\varepsilon ({\bf k})$,
the areas under the peak $W_p({\bf k})$ and $W_h({\bf k})$ and the imaginary
part of the self-energy
for $J/\tau =0.2$, $h/\tau=0.3$
and ${\bf k}=n(\pi /20,\pi /20)$ }
\label{t1}

\begin{tabular}{ccccc}
$n$ & $\varepsilon ({\bf k})$ & $W_p({\bf k})$  & $W_h({\bf k})$ &
 $-Im\Sigma ({\bf k},\varepsilon ({\bf k}))$  \\
\hline
$0 \ast    $ &   -4.30    &  0.020 & 0      &   0.1647 \\
$1 \ast    $ &   -4.31    &  0.018 & 0.000  &   0.1751 \\
$2 \ast    $ &   -4.35    &  0.008 & 0.001  &   0.1136 \\
$3 \ast    $ &   -4.40    &  0.008 & 0.001  &   0.0999 \\
$4 \ast    $ &   -4.47    &  0.010 & 0.002  &   0.0798 \\
$5 \ast    $ &   -4.55    &  0.022 & 0.004  &   0.0472 \\
$6 \ast    $ &   -4.64    &  0.061 & 0.013  &   0.0259 \\
$7 \circ   $ &   -4.74    &  0.718 & 0.170  &   0.0020 \\
$8 \bullet $ &   -4.84    &  0.849 & 0.210  &   0.0000 \\
$9 \bullet $ &   -4.89    &  0.850 & 0.218  &   0.0000 \\
$10 \bullet$ &   -4.90    &  0.878 & 0.231  &   0.0000 \\
$11 \bullet$ &   -4.86    &  0.785 & 0.212  &   0.0000 \\
$12 \ast   $ &   -4.80    &  0.232 & 0.064  &   0.0148 \\
$13 \ast   $ &   -4.74    &  0.059 & 0.017  &   0.0445 \\
$14 \circ  $ &   -4.65    &  0.100 & 0.029  &   0.0251 \\
$15 \circ  $ &   -4.57    &  0.095 & 0.029  &   0.0265 \\
$16 \circ  $ &   -4.50    &  0.116 & 0.037  &   0.0086 \\
$17 \circ  $ &   -4.43    &  0.132 & 0.044  &   0.0044 \\
$18 \circ  $ &   -4.39    &  0.145 & 0.051  &   0.0012 \\
$19 \circ  $ &   -4.36    &  0.148 & 0.053  &   0.0003 \\
$20 \circ  $ &   -4.36    &  0.146 & 0.053  &   0.0001 \\

\end{tabular}

\end{table}

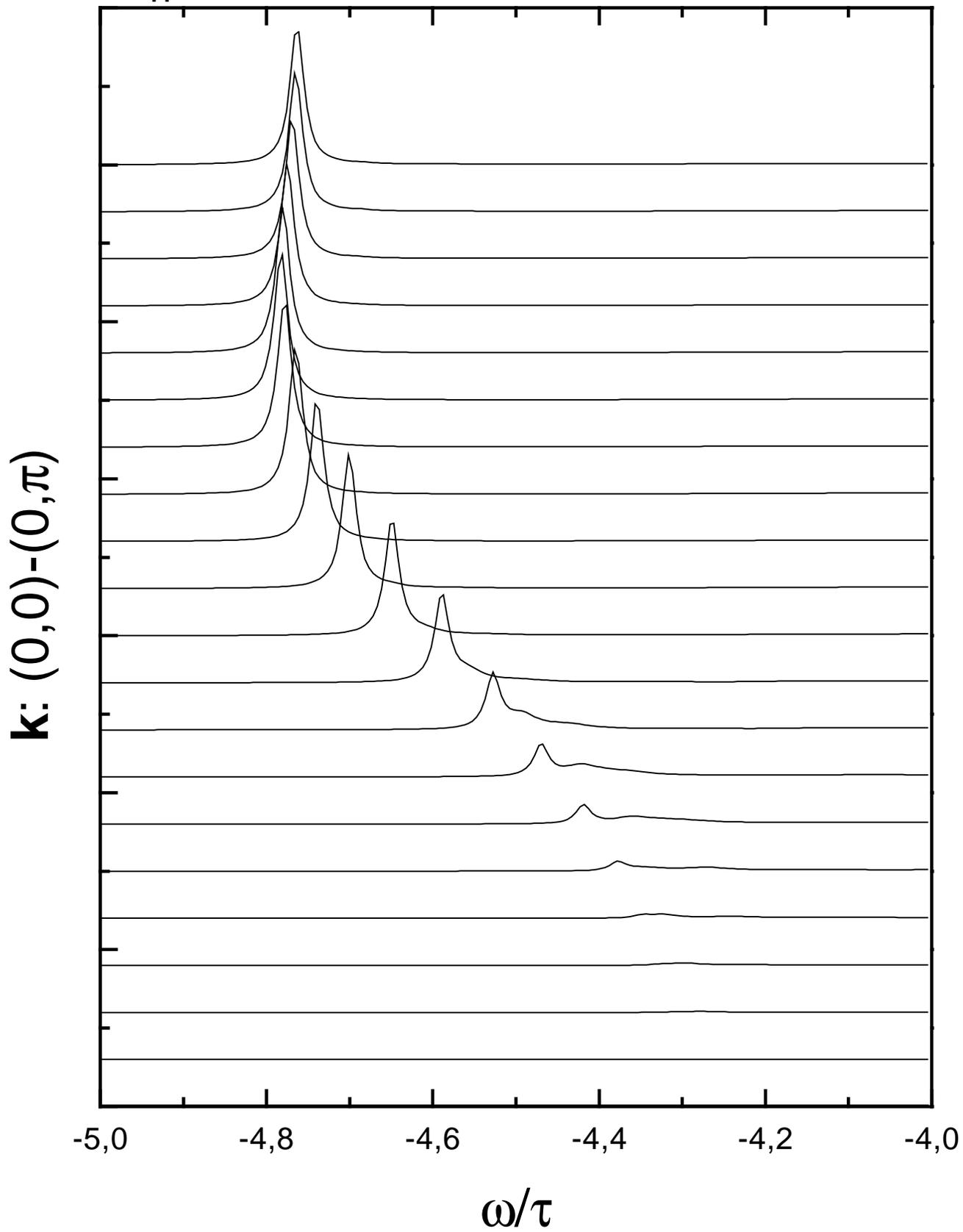
\end{document}